\documentclass{article}   
\pdfoutput=1

\usepackage{comment}
\usepackage{amsfonts}
\usepackage{amsmath}
\usepackage{amsthm}
\usepackage{graphicx} 
\usepackage{caption}
\usepackage{subcaption}
\usepackage{algpseudocode}
\usepackage{algorithmicx}
\usepackage{algorithm}
\usepackage{url}       
\usepackage{tabu}
\usepackage{fullpage}

                               \newcommand{\E}{\mathbb{E}}

                               \newcommand{\I}{\mathcal{I}}
                               
                               \newcommand{\R}{\mathbb{R}}

                               \newcommand{\tr}{\operatorname{tr}}

\newtheorem{prop}{Proposition}

\begin{document}

\title{A note on optimal experiment design for nonlinear systems using dynamic programming}

\author{John Maidens and Murat Arcak \thanks{Department of Electrical Engineering \& Computer Sciences,
University of California, Berkeley,
Cory Hall, Berkeley, CA, 94720, USA. 
\texttt{\{maidens, arcak\}@eecs.berkeley.edu}}
 }

\maketitle

\section{Main result} 
We consider a discrete-time dynamical system with noisy observations
\begin{equation}
\begin{split}
    x_{t +1}   &=        f(x_t, u_t, \theta) \\
    Y_t          &\sim   p_{x_t}(y_t )
\end{split} 
\label{eq:system}
\end{equation}
where $x_t \in \R^n$ denotes the system's state, $u_t \in \R^m$ is a sequence of inputs to be designed and $\theta \in \R^p$ is a vector of unknown parameters that we wish to estimate. Observations are drawn independently from a known distribution that is parametrized by the system state $x_t$. We assume that for all $x_t \in \R^n$ the probability distribution is absolutely continuous with respect to some measure $\mu$ and we denote its density with respect to $\mu$ by $p_{x_t}(y_t)$. We further assume that this density is differentiable with respect to the parameter $x_t$ and define the Fisher information matrix as
\[
I _{Y_t}(x_t) = \E_\theta \left[ \Big( \nabla_{x_t} \log p_{x_t}(Y_t) \Big)  \Big( \nabla_{x_t} \log p_{x_t}(Y_t) \Big)^T 
    \right].
\]

We consider this system over a finite horizon $0 \le t \le N$. Our goal is to design a sequence $u$ that provides a maximal amount of information about the unknown parameter vector $\theta$ for $\theta$ in a neighbourhood of some nominal value of the parameters $\theta_0$. Mathematically, we wish to choose $u$ to maximize a function $\phi(\I _Y (\theta_0))$ of the Fisher information that the joint output $Y = (Y_0, \dots, Y_N)$ carries about the parameter $\theta$. The function $\phi$ is chosen to be a measure of the ``largeness'' of the positive semidefinite matrix $\I $. Multiple choices for the function $\phi$ have been proposed \cite{Pukelsheim06}; here we use $\phi(\I ) = \tr(\I)$ (often called ``$T$-optimal design''). In general this problem is nonconvex as a function of $u$, but due to the fact that the trace is linear our objective function is additive so a global solution can be found using dynamic programming. 

The following result allows us to compute the information contained in the observed data. A proof of this proposition is given in Section \ref{sec:proof}. 
\begin{prop}
	Suppose that for all $x_t \in \R^n$ the density $p_{x_t}(y_t)$ is differentiable with respect to $x_t$ and that there exists a $\mu$-integrable function $q$ with $\left| \frac{\partial p_{x_t}(y_t) }{\partial x_t} \right| \le q(y_t)$ for all $y_t \in \R$.  If $f$ is $C^1$ in $x_t$ and $\theta$, then the Fisher information with respect to the parameter $\theta$ can be computed as 
	\begin{equation}
		\I _Y(\theta) = \sum_{t=0}^N (\nabla_\theta x_t) ^T \I _{Y_t}(x_t) (\nabla_\theta x_t)
	\end{equation} 
	where $\nabla_\theta x_t$ denotes the Jacobian of $x_t$ with respect to $\theta$.
\label{prop:Fisher} 
\end{prop} 

Thus, the $T$-optimal design criterion, the objective function is given by
\begin{equation}
	  \tr(\I _Y(\theta)) = \sum_{t=0}^N \tr\Big( (\nabla_\theta x_t) ^T \I _{Y_t}(x_t) (\nabla_\theta x_t) \Big).  
\label{eq:cost}
\end{equation} 
Applying the chain rule to (\ref{eq:system}), we get a dynamical system 
\begin{equation}
 	\nabla_\theta x_{t+1} = \nabla_\theta f( x_t, u_t, \theta) + \nabla_x f(x_t, u_t, \theta) \ \nabla_\theta x_t
  \label{eq:matrix_system}
\end{equation}
describing the time evolution of the sensitivities $\nabla_\theta x_t$. The dynamics (\ref{eq:system}) and (\ref{eq:matrix_system}) together with the cost function (\ref{eq:cost}) define a discrete-time finite-horizon optimal control problem that can be solved via dynamic programming \cite{Bertsekas95}. In particular, we define the sequence of value functions $J_k$ via
\begin{equation*}
\begin{split}
 	J_N(x_N, \nabla_\theta x_N) &= \tr\Big( (\nabla_\theta x_N) ^T \I _{Y_N}(x_N) (\nabla_\theta x_N) \Big)  \\
 	J_k(x_k,  \nabla_\theta x_k)  &= \max_{u_k}  \left\{  \tr \Big( (\nabla_\theta x_k) ^T \I _{Y_k}(x_k) (\nabla_\theta x_k) \Big) 		+ J_{k+1} \Big( f(x_k, u_k, \theta),  \nabla_\theta f( x_k, u_k, \theta) + \nabla_x f(x_k, u_k, \theta) \ \nabla_\theta x_k \Big) \right\}.
\end{split}
\label{eq:DP}
\end{equation*}
If the control policy $u^*_k = \mu_k^*(x_k, \nabla_\theta x_k)$ maximizes the right hand side of (\ref{eq:DP}) then $u^*$ is globally optimal.

Related approaches to the optimal experiment design problem appear in \cite{Chen75} and \cite{Morelli90} for continuous-time dynamical systems with Gaussian noise. However, a different objective function $\phi$ is used and these approaches requires appending a nonlinear matrix differential equation for the dispersion (the inverse of the Fisher information) to the system state in addition to equation (\ref{eq:matrix_system}). By choosing the $T$-optimal design criterion $\phi(\I) = \tr(\I )$, we are able to avoid adding an equation for the dispersion to the system state, allowing us to efficiently solve problems of larger dimension.

\section{Example problem} 

We consider a population of fruit flies, whose dynamics are modelled using the discrete logistic equation 
\[
 	x_{t+1} = x_t + r x_t (K - x_t).
\]
We want to estimate the reproduction rate $r$ along with the carrying capacity $K$. To generate data from which to estimate the model parameters, we place a sequence of traps into the fly cage, each capturing a fraction $u_t$ of the current fly population. By measuring the number of fruit flies caught in the trap, we wish to infer the model parameters. The optimization problem thus consists of choosing the size of the traps (and hence the proportion of flies trapped) at each sampling interval. 

This leads to a model for the population dynamics together with the number of fruit flies trapped $Y_t$ 
\begin{equation}
\begin{split}
  x_0       &= K \\
  x_{t+1}  &= x_t(1 - u_t) + r x_t(1-u_t) (K - x_t(1-u_t)) \\
  Y_t        &\sim Poisson(x_t u_t) . 
\end{split} 
\label{eq:flies}
\end{equation} 

\begin{figure}[ht!]
  \centering 
  \begin{subfigure}[b]{0.4\textwidth}
	  \includegraphics[width=\textwidth]{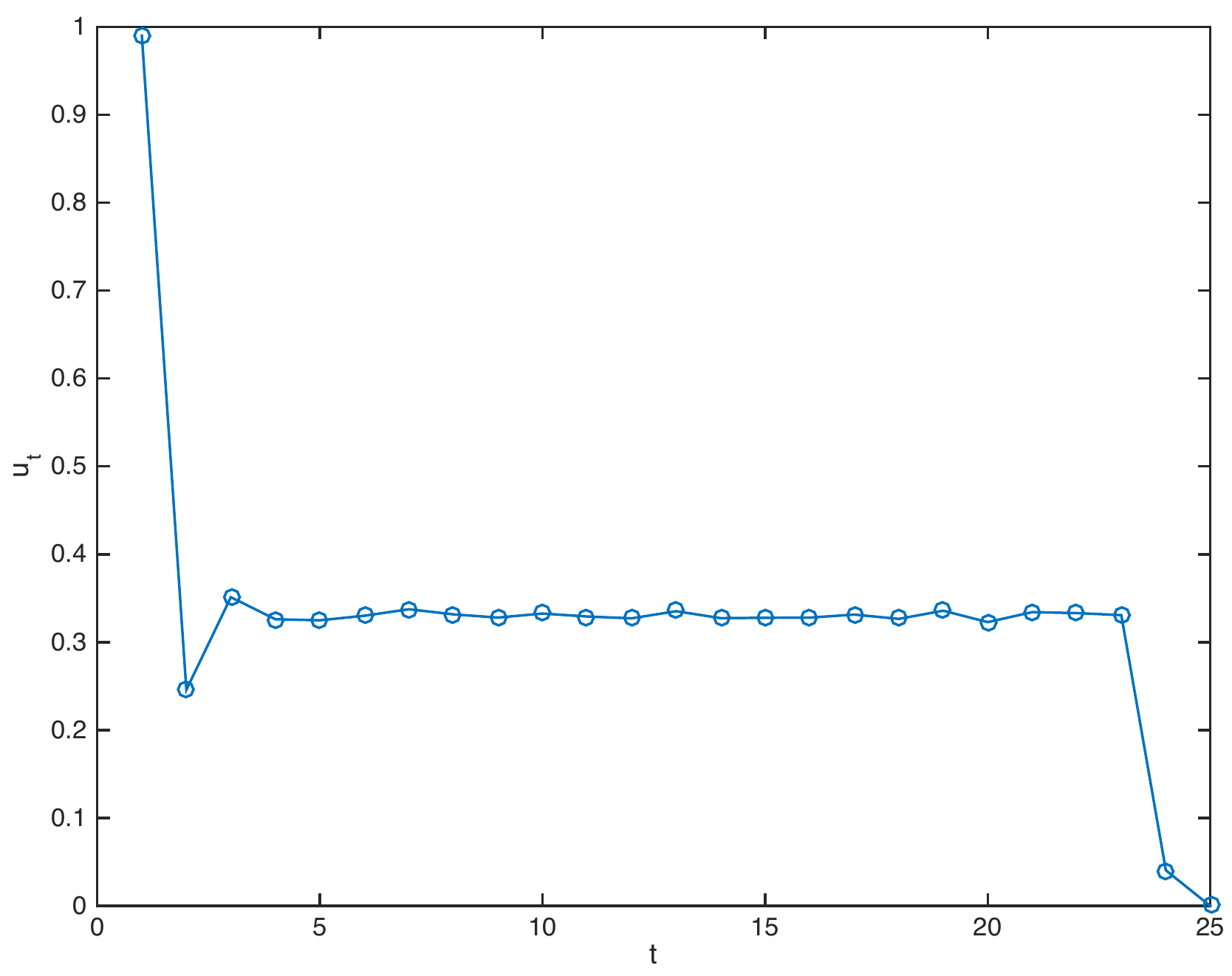}
	  \caption{Optimal input trajectory for (\ref{eq:flies}) computed using dynamic programming} 
	  \label{fig:optimal_inputs}
   \end{subfigure} 
   ~
  \begin{subfigure}[b]{0.4\textwidth}
	  \includegraphics[width=\textwidth]{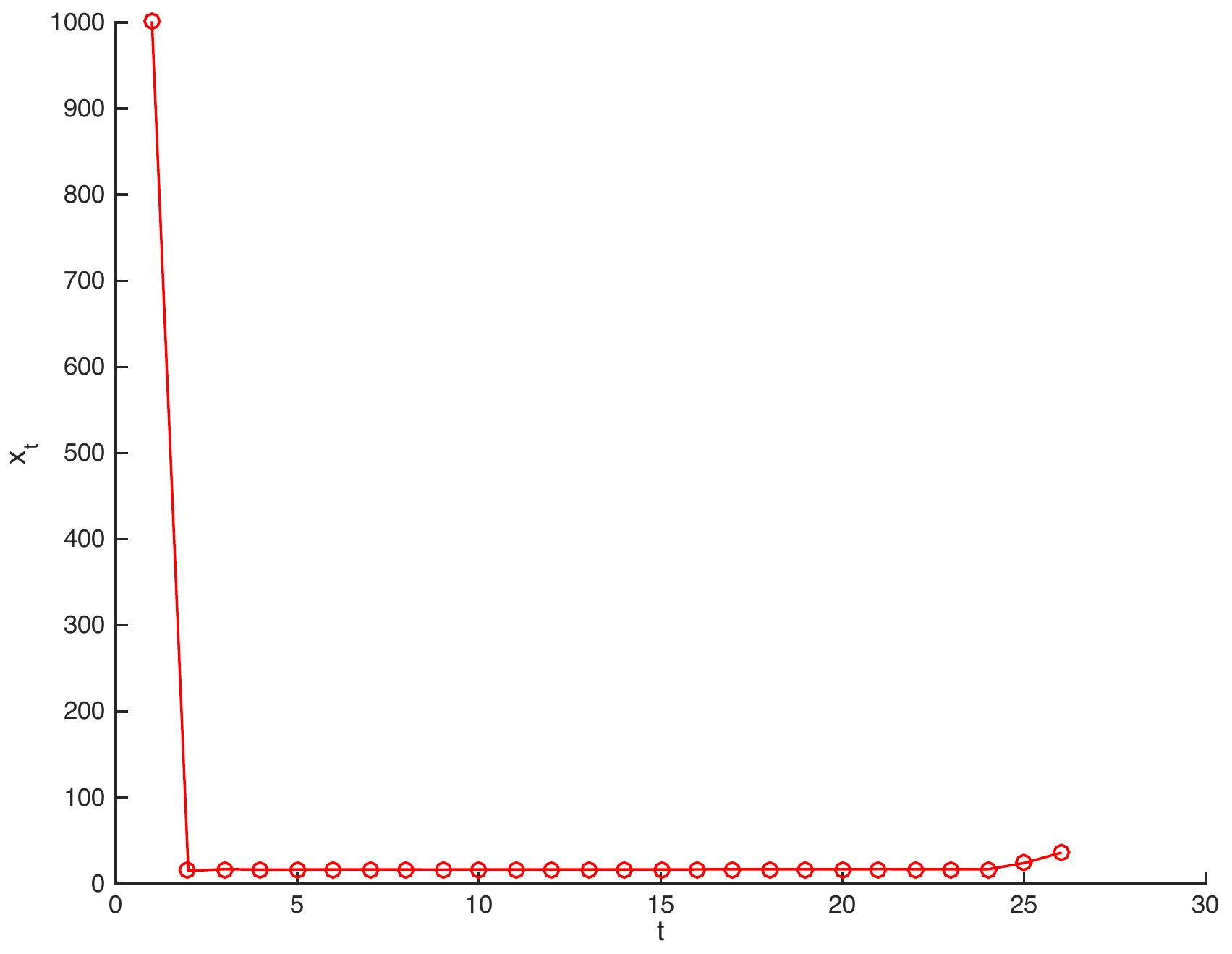}
	  \caption{State trajectory corresponding to the input given in Figure \ref{fig:optimal_inputs}} 
	  \label{fig:optimal_states} 
  \end{subfigure}
  \caption{Numerically-computed solution to the optimal experiment design problem}
\end{figure}

For this problem, we approximate the functions $J_k$ by evaluation on a grid of size $100 \times 100 \times 100$. We optimize about the nominal parameter values $r_0 = 5 \times 10^{-4}$ and $K_0 = 1000$. This is implemented in MATLAB using the dynamic programming routine introduced in \cite{Sundstrom09}. The optimal inputs are computed in 33.01 seconds and are shown in Figure \ref{fig:optimal_inputs}. The corresponding state trajectory is shown in Figure \ref{fig:optimal_states}. 

We see that the optimal observation scheme is to first capture a large fraction $u_1 = 0.9795$ of the flies allowing us to get a reliable estimate for the carrying capacity $K$. After this, we capture a fraction $u_t \approx 0.32$ of the flies, just enough to keep the population constant. This provides maximal sensitivity to the growth rate $r$ in a neighbourhood of $r_0$. Indeed, if $r > r_0$ we will see the the population of flies grow over time, whereas if $r < r_0$ the population will shrink toward zero.

\section{Proof of Proposition \ref{prop:Fisher}}
\label{sec:proof} 

\begin{proof}
  First, note that the hypotheses of this proposition provide sufficient regularity to exchange the order of differentiation with respect to $\theta_i$ and integration with respect to $y_t$. Therefore for all $i = 1, \dots, p$ and $t = 0, \dots, N$ 
  \begin{equation*}
  \begin{split}
    \E_\theta \left[ \frac{\partial \log p_\theta(Y_t)}{\partial \theta_i} \right]  
    &= \int \frac{\partial \log p_\theta (y_t)}{\partial \theta_i} p_\theta(y_t) d\mu(y_t) 
    = \int \frac{\partial p_\theta (y_t)}{\partial \theta_i} d\mu(y_t) \\ 
    &= \frac{\partial}{\partial \theta_i} \int p_\theta(y_t) d\mu(y_t)  =  \frac{\partial}{\partial \theta_i} 1 = 0. 
      \end{split}
  \end{equation*} 
  Now for all $i, j = 1, \dots, p$ we can compute the $(i, j)$-th entry of $\I _\theta(Y)$ as 
  \begin{equation*}
  \begin{split}
    \I _Y(\theta)_{i, j} 
    &= \E_\theta \left[ \frac{\partial \log p_\theta(Y)}{\partial \theta_i} \frac{\partial \log p_\theta(Y)}{\partial \theta_j}\right] \\
    &= \E_\theta \left[\left( \sum_{t=0}^N \frac{\partial \log p_\theta(Y_t)}{\partial \theta_i} \right) \left( \sum_{s=0}^N \frac{\partial \log p_\theta(Y_s)}{\partial \theta_j} \right) \right] \\ 
    &= \sum_{t=0}^N  \E_\theta \left[ \frac{\partial \log p_\theta(Y_t)}{\partial \theta_i} \frac{\partial \log p_\theta(Y_t)}{\partial \theta_j}\right] + \sum_{t=0}^N \sum_{s = 0,  s\ne t}^N   \E_\theta \left[ \frac{\partial \log p_\theta(Y_t)}{\partial \theta_i} \right]    \E_\theta \left[ \frac{\partial \log p_\theta(Y_s)}{\partial \theta_j} \right]  \\
    &= \sum_{t=0}^N  \E_\theta \left[ \frac{\partial \log p_{x_t(\theta)} (Y_t)}{\partial \theta_i} \frac{\partial \log p_{x_t(\theta)}(Y_t)}{\partial \theta_j}\right] \\
    &= \sum_{t=0}^N  \E_\theta \left[ \langle \frac{\partial x_t}{\partial \theta_i}, \nabla_{x_t} \log p_{x_t}(Y_t) \rangle \langle \frac{\partial x_t}{\partial \theta_j}, \nabla_{x_t} \log p_{x_t}(Y_t) \rangle \right] \\ 
    &= \sum_{t=0}^N \frac{\partial x_t}{\partial \theta_i}^T \E_\theta \left[ \Big( \nabla_{x_t} \log p_{x_t}(Y_t) \Big)  \Big( \nabla_{x_t} \log p_{x_t}(Y_t) \Big)^T 
    \right]  \frac{\partial x_t}{\partial \theta_y}  \\
    &=  \sum_{t=0}^N \frac{\partial x_t}{\partial \theta_i}^T \I _{Y_t}(x_t) \frac{\partial x_t}{\partial \theta_y}. 
  \end{split}
  \end{equation*} 
  So 
  \[
    \I _Y(\theta) = \sum_{t=0}^N (\nabla_\theta x_t) ^T \I _{Y_t}(x_t) (\nabla_\theta x_t). 
  \]
\end{proof}

\bibliographystyle{ieeetran}
\bibliography{optimal_design} 

\begin{thebibliography}{1}
\providecommand{\url}[1]{#1}
\csname url@samestyle\endcsname
\providecommand{\newblock}{\relax}
\providecommand{\bibinfo}[2]{#2}
\providecommand{\BIBentrySTDinterwordspacing}{\spaceskip=0pt\relax}
\providecommand{\BIBentryALTinterwordstretchfactor}{4}
\providecommand{\BIBentryALTinterwordspacing}{\spaceskip=\fontdimen2\font plus
\BIBentryALTinterwordstretchfactor\fontdimen3\font minus
  \fontdimen4\font\relax}
\providecommand{\BIBforeignlanguage}[2]{{%
\expandafter\ifx\csname l@#1\endcsname\relax
\typeout{** WARNING: IEEEtran.bst: No hyphenation pattern has been}%
\typeout{** loaded for the language `#1'. Using the pattern for}%
\typeout{** the default language instead.}%
\else
\language=\csname l@#1\endcsname
\fi
#2}}
\providecommand{\BIBdecl}{\relax}
\BIBdecl

\bibitem{Pukelsheim06}
F.~Pukelsheim, \emph{Optimal Design of Experiments}.\hskip 1em plus 0.5em minus
  0.4em\relax Society for Industrial and Applied Mathematics, 2006.

\bibitem{Bertsekas95}
D.~P. Bertsekas, \emph{Dynamic Programming and Optimal Control, Volume
  1}.\hskip 1em plus 0.5em minus 0.4em\relax Athena Scientific, 1995.

\bibitem{Chen75}
R.~T.~N. Chen, ``Input design for aircraft parameter identification: Using
  time-optimal control formulation,'' in \emph{Methods for Aircraft State and
  Parameter Identification, Advisory Group for Aerospace Research and
  Development (AGARD), Conference Proceedings no. 172}, 1975.

\bibitem{Morelli90}
E.~A. Morelli and V.~Klein, ``Optimal input design for aircraft parameter
  estimation using dynamic programming principles,'' in \emph{AIAA Atmospheric
  Flight Mechanics Conference paper 90-2801}, 1990.

\bibitem{Sundstrom09}
O.~Sundstr\"{o}m and L.~Guzzella, ``A generic dynamic programming {M}atlab
  function,'' in \emph{IEEE International Symposium on Control Applications and
  Intelligent Control (CCA \& ISIC)}, 2009, pp. 1625--1630.

\end{thebibliography}

\end{document}